\documentclass[runningheads]{llncs}
\usepackage{graphicx}

\usepackage{hyperref}

\hypersetup{
  colorlinks   = true,    
  urlcolor     = blue,    
  linkcolor    = blue,    
  citecolor    = blue      
}
\usepackage[table,xcdraw]{xcolor}
\usepackage{color}

\usepackage[colorinlistoftodos]{todonotes}
\usepackage[inline]{enumitem}
\usepackage{multirow}
\usepackage[normalem]{ulem}
\usepackage{booktabs}
\usepackage{soul,color}
\usepackage{ctable}
\usepackage{cite}
\usepackage{makecell}
\usepackage{array}

\usepackage{booktabs,siunitx}

\usepackage{xcolor,colortbl}
\definecolor{softgrey}{rgb}{0.9, 0.9, 0.9}

\soulregister\cite7
\soulregister\ref7
\soulregister\pageref7

\useunder{\uline}{\ul}{}

\begin{document}
\title{Requirements Quality vs Process and Stakeholders' Well-being: A Case of a Nordic Bank}
\titlerunning{Requirements Quality vs Process and Stakeholders' Well-being}
%
\author{Emil Lind \inst{1}
\and
Javier Gonzalez-Huerta\inst{1}\orcidID{0000-0003-1350-7030} \and
Emil Al\'egroth\inst{1}\orcidID{0000-0001-7526-3727}}
\authorrunning{Lind, Gonzalez-Huerta, and Al\'egroth}
%
\institute{
Software Engineering Research Lab SERL, Blekinge Institute of Technology,\\ 371 79, Karlskrona, Sweden \\
\email{emil.lind@bth.se},
\email{javier.gonzalez.huerta@bth.se},
\email{emil.alegroth@bth.se}}
\maketitle              

\begin{abstract}~Requirements are key artefacts to describe the intended purpose of a software system. The quality of requirements is crucial for deciding what to do next, impacting the development process's effectiveness and efficiency. However, we know very little about the connection between practitioners' perceptions regarding requirements quality and its impact on the process or the feelings of the professionals involved in the development process.

\noindent \textbf{Objectives:} This study investigates: i) How software development practitioners define requirements quality, ii) how the perceived quality of requirements impact process and stakeholders' well-being, and iii) what are the causes and potential solutions for poor-quality requirements.

\noindent \textbf{Method:} This study was performed as a descriptive interview study at a sub-organization of a Nordic bank that develops its own web and mobile apps. The data collection comprises interviews with 20 practitioners, including requirements engineers, developers, testers, and newly employed developers, with five interviewees from each group.

\noindent \textbf{Results:} The results show that different roles have different views on what makes a requirement good quality. Participants highlighted that, in general, they experience negative emotions, more work, and overhead communication when they work with requirements they perceive to be of poor quality. The practitioners also describe positive effects on their performance and positive feelings when they work with requirements that they perceive to be good.

\end{abstract}

\keywords{Requirements  Engineering\and Requirements Quality \and Human Factors \and Empirical Study}

\section{Introduction}\label{sec:intro}

Requirements are crucial for developing software-intensive products and services since they are the main link between the business value and its implementation. 
As such, the consequences of issues---Poor quality such as incompleteness or ambiguity---with requirements might lead to a project or product failure~\cite{Kamata,MendezNapire,Fernandez2017}. 
Requirements are used by multiple roles, including developers, testers, and user experience designers in their daily work~\cite{Femmer2019}. 
Therefore, requirements quality has a profound, direct impact on the outcome of the different downstream activities in the development process and on the quality of the final product itself~\cite{Femmer2019}.

Moreover, changes to requirements have an intrinsic relationship to project failure and results in projects not being finished within time or budget constraints~\cite{Zowghi2002}. 
Changes to requirements, before and after release, affect the different development activities~\cite{Javed2004}, for instance, by forcing the re-prioritization of tasks and effort allocation.

Several standards define how to write good requirements (e.g., ISO29148~\cite{ISO29148} or IREB \cite{IREB}) and have also been studied in several research works (e.g., \cite{Femmer2019,Fernandez2017}).
These works aim to provide an objective, general view of how a good quality requirement should be. 
However, from a practitioner's view, there is a lack of understanding regarding what they perceive as good - or bad- requirements and how they affect their daily work. 
Following Femmer's and Vogelsang's activity-based view on requirements and their quality~\cite{Femmer2019}, it is highly relevant to identify the practitioners' view on requirements quality and how practitioners subjectively define it.
The reason is that standards are often too general or imprecise to be applied in different industries. 
Following this reasoning, eliciting developers', testers', and requirements engineers' experiences and how they are affected by what they perceive to be bad requirements - compared to what they perceive to be good requirements - is therefore of importance. The reason is that the practitioners' needs may not align with what is prioritized in the standards. 
Thus, research into the phenomenon provides insights into these practitioners' ways of working and inputs for future improvements to said requirements standards.

There are research works that analyze the impact of good/bad requirements on the project outcomes (e.g., \cite{MendezNapire,Fernandez2017,Kamata,Rempel2017,Damian2006}). However, these are either based on questionnaire surveys or directly based on static analysis techniques. Thereby leaving a gap in knowledge from empirical case studies that go deeper, through interviews and focus groups, to understand the consequences of good and bad quality requirements as the practitioners perceive them.

This study investigates the differences in how practitioners from different roles define good and bad requirements, i.e., what characteristics make requirements good quality. 
Additionally, the study aims to determine the impact practitioners experience from good or bad quality requirements in their work, workload, and well-being.
Furthermore, the study also aims to find the perceived causes and potential solutions to poor quality requirements. The goal is also to gain an understanding of requirements quality, which is essential first to align with existing standards but also to understand what are good-enough requirements that allow organizations to prioritize requirements for implementation that add value to the product~\cite{Ernst2021}.

The remainder of the paper is structured as follows: Section \ref{sec:related} discusses related research in the area. Section~\ref{sec:methodology} describes the researfch methodology followed in the interview study. Section~\ref{sections:results} reports the main results of the study. Section~\ref{sec:discussion} discusses the main findings. In Section~\ref{sections:threats} we discuss the limitations and threats to the validity. Finally, Section~\ref{sec:conclusions} draws the main conclusions and discusses further works.
\section{Related Work}\label{sec:related}

Requirements Engineering (RE) in general and specific RE methods are well represented in the body of scientific knowledge. 
There are also recommendations and guidelines for working with RE and even quality standards for requirements (e.g., \cite{ISO29148,IREB}).

The NaPiRE~\cite{MendezNapire,Fernandez2017} project, which involves more than 200 companies in 10 countries, has mapped several kinds of bad requirements with factors for project failure or linked these requirements problems with project delays or budget overruns. 
Similarly, several studies (e.g., \cite{Damian2006,Kamata,Zowghi2002,Javed2004}) has tried to find relationships between requirements quality to requirements (i.e., requirements smells~\cite{Femmer2017}). 
However, what is still unclear is what the impact of these smells would be.
Femmer and Vogelsang~\cite{Femmer2019} found a relationship between the quality of requirements and the \textit{quality in use} of the software system being developed.

However, neither the NaPiRE project nor the studies mentioned above have considered the different perceptions of what good or bad requirements are for different roles or the effect that bad requirements might have on the practitioners' work, workload and well-being. Well-being, specially stress has been found as an important factor for hindering collaborative work and technical practices~\cite{Meier2018}.
All these aspects are essential to define good-enough requirements that allow organizations to prioritize requirements for implementation that add value to the product~\cite{Ernst2021}, thus motivating their study in the area of RE.

\section{Research Methodology}\label{sec:methodology}

The study addressed the following research questions:
\begin{itemize}
	\item \textbf{RQ1} How do software development practitioners define requirements quality?
	\item \textbf{RQ2} How does the perceived quality of requirements impact the work of practitioners in software development?
	\item \textbf{RQ3} What are the perceived causes and potential solutions of the poor quality of requirements?
\end{itemize}

The intent is to map the effects of low-quality requirements (as per RQ1), as perceived by requirements engineers, developers, and testers, to the enjoyment of work, stress, and well-being in general, as the perception of their colleagues, organization, and workload. Hence, the results lean into the human factors domain of software engineering or behavioural software engineering\cite{Lenberg2014}. Such results, albeit often less tangible than technical factors, are essential for the general understanding of software engineering.

\subsection{Context, Case, and Unit of Analysis}\label{sec:case}

To address the research questions mentioned above,  we conducted an interview study in an industrial setting comprising twenty interviews with software development practitioners. 
We conducted the study in a sub-organization of a Nordic bank\footnote{Fictitious name to preserve anonymity} which develops mobile and web apps for the bank's end-users, therefore belonging to the financial technology (fintech) domain.

\begin{figure}[tb!]
\includegraphics[width=\textwidth]{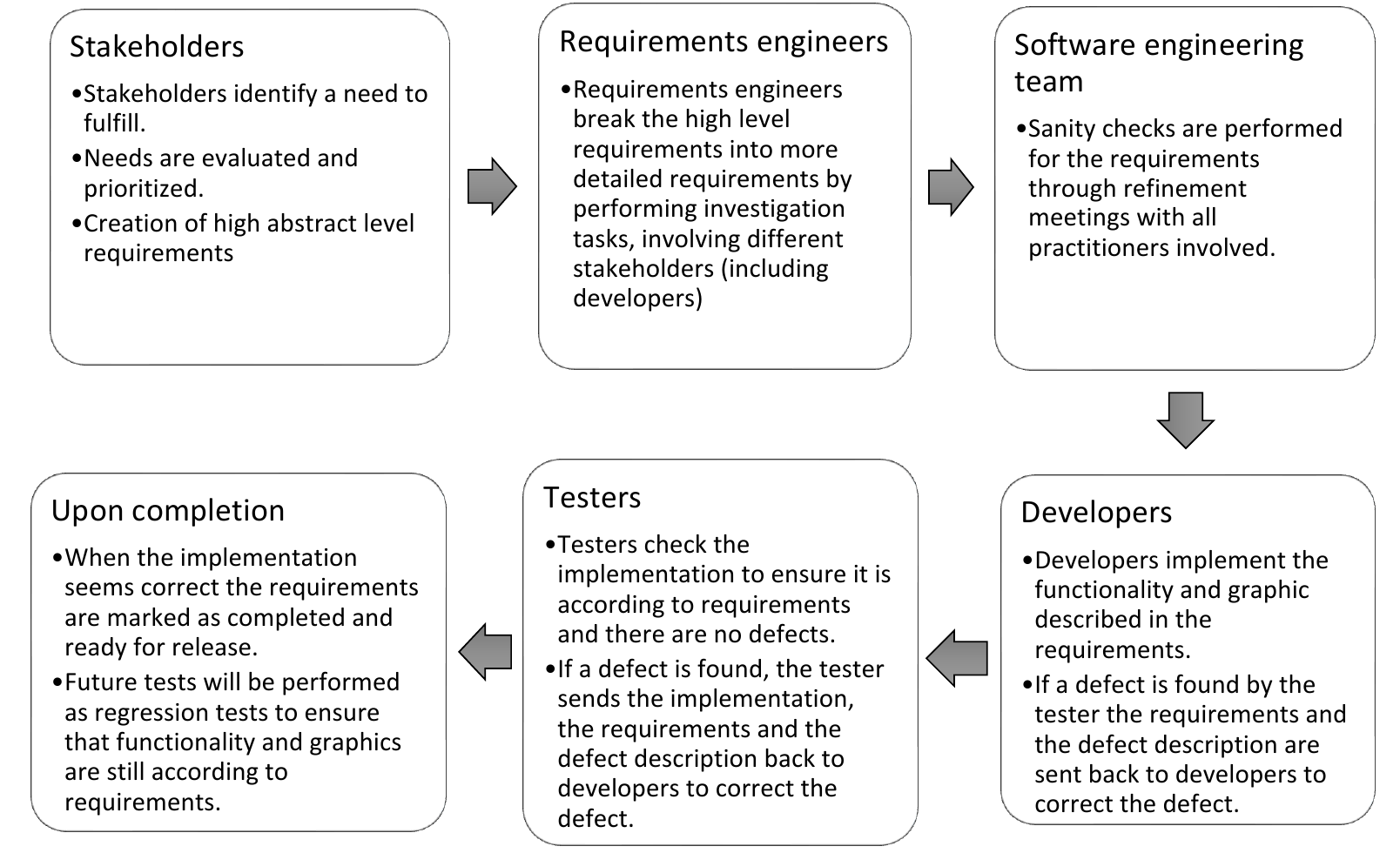}
\caption{Requirements work-process-related codes} \label{fig:workingwithreqs}
\end{figure}

The organization has development teams organized into different areas, such as product areas or business/domain areas, referred to as value streams. On average, the development teams consist of 10 to 12 employees and contain roles such as requirements engineers, testers, developers, UX designers, and scrum masters. 
Note that the studied organization carries out requirements engineering, development and software testing in a downstream process, i.e., there is a hand-over between requirements engineers to developers and later a hand-over from developers to testers. Fig.~\ref{fig:workingwithreqs} provides a visual, brief overview of the development workflow used at the organization.

In this study, we focus on four different \textit{units of analysis}, i.e., requirements engineers, testers, software developers with more than one year of experience at the organization and recently recruited software developers.
Employees from each group were selected, by a hybrid of convenience and random sampling~\cite{Linaker2015}, from different projects and geographic locations to acquire a more representative sample inside the organization. The hybrid sampling consisted of using a list of employees with suitable characteristics for the study provided by managers at the organization.
We refer to this as hybrid sampling since the authors had limited control over which participants were added to the list, i.e. the participants' managers ultimately were the ones suggesting their participation in the study. We selected the organization also by convenience since it is one of the partner companies in an ongoing research project that focuses, among other topics, on addressing the quality degradation of software assets. Table~\ref{table:demographics} details the participants' demographic information.

\begin{table}[tb]
\begin{center}
\scriptsize
\caption{Demographic information of the interview participants, including interviewee Id, participant's working location, and years of experience within the organisation} \label{table:demographics} <
\begin{tabular}{p{0.5\textwidth}p{0.15\textwidth}p{0.15\textwidth}}
\hline
\textbf{Interviewee Id (including role)} & \textbf{Location}       & \textbf{Experience}\\
\hline
Requirements Engineering 1      & Sweden         & \textless{}5 years                     \\
Requirements Engineering 2      & Sweden         & \textgreater{}5 years                  \\
Requirements Engineering 3      & Baltic Country & \textgreater{}5 years                  \\
Requirements Engineering 4      & Sweden         & \textless{}5 years                     \\
Requirements Engineering 5      & Sweden         & \textgreater{}5 years                  \\
Developer 1                     & Sweden         & \textless{}5 years                     \\
Developer 2                     & Sweden         & \textless{}5 years                     \\
Developer 3                     & Sweden         & \textgreater{}5 years                  \\
Developer 4                     & Sweden         & \textless{}5 years                     \\
Developer 5                     & Sweden         & \textless{}5 years                     \\
Tester 1                        & Sweden         & \textgreater{}5 years                  \\
Tester 2                        & Baltic Country & \textless{}5 years                     \\
Tester 3                        & Baltic Country & \textless{}5 years                     \\
Tester 4                        & Sweden         & \textless{}5 years                     \\
Tester 5                        & Baltic Country & \textless{}5 years                     \\
Recently recruited developer 1 & Sweden         & \textless{}3 months                    \\
Recently recruited developer 2 & Sweden         & \textless{}3 months                    \\
Recently recruited developer 3 & Sweden         & \textless{}3 months                    \\
Recently recruited developer 4 & Sweden         & \textless{}3 months                    \\
Recently recruited developer 5 & Sweden         & \textless{}3 months                    \\\hline
\end{tabular}

\end{center}
\end{table}

The recently recruited developers were an opportunity-based unit of analysis, interviewed to complement the results from the study's three central units of analysis (i.e., requirements engineers, testers, and developers). 
We sampled these participants following the same approach as the other participants. 
However, the sample frame of potential participants was much smaller, i.e. only employees with less than four months of employment were eligible.
However, since we exercised no control over the selection, we still classify it as hybrid convenience and random selection.

In total, we conducted interviews with $20$ participants; five requirements engineers, five developers, five testers, and five recently recruited developers. 
We interviewed the recently recruited developers twice, the first time when they had finished or were close to finishing their onboarding at the organization and the second time when they had worked for a few more months at the organization\footnote{Although the analysis of the differences between these two interview instances is out of the scope of this paper.} Therefore the total number of interviews conducted in the study was $25$. 

To ensure anonymity, we clustered the participants' experience into groups of more than five years, less than five years (the requirement for participating was at least a year in the organization), and three months or less for the recently recruited developers.

\subsection{Data Collection}

Data for this interview study was collected using semi-structured interviews. The first part of the interview guide aimed at answering RQ1, whilst the second part of the interview aimed at answering RQ2 and RQ3. 

Each interview took thirty to sixty minutes, following a predefined interview guide\footnote{The interview guide is available in the companion materials in Zenodo DOI:\href{https://doi.org/10.5281/zenodo.7306032}{10.5281/zenodo.7306032}}, recorded with audio and video, and later transcribed to text. 
The interview guide consisted of 16 predefined questions for the testers and developers that had worked in the organization for at least one year. For requirements engineers, the interview guide consisted of 24 predefined questions, 19 predefined questions for the first interview with the recently recruited developers and 15 predefined questions for the second interview with the recently recruited developers. 
Although the interview guides varied depending on the interviewees' roles, the semantic information gathered aimed at providing complementing answers to the research questions. 
The guides also had questions that are not mapped to any specific research question. We added these additional questions to gather supplementary information to understand the context and to interpret the interview results that contributed to the research questions. 
The number of predefined questions was decided to give enough time to ask follow-up questions.

\subsection{Data Analysis}

The interviews were analyzed using thematic analysis\cite{Cruzes2011,Braun2006}.
Open coding was used, where codes were generated based on the semantic meaning of statements from the interview transcripts, using mainly a deductive approach~\cite{Cruzes2011}. 
We used the coded information and the associated quotes to synthesize evidence from the collected data. This evidence-driven analysis approach was suitable for answering the research questions due to the study's descriptive nature.

We added the codes incrementally from the interview results. As stated, the codes were formulated based on the semantic meaning of the interviewees' statements. When another statement was found to contain similar semantic information, said the statement was marked with the same code. We did not restrict coding to a 1-to-1 mapping between codes and statements. Hence, we could code a statement with one or several codes. We stored all extracted statements from the transcripts with the codes in the code books for consistency.

The rationale for using coding was to provide an overview of the data to connect statements and observations to draw higher-level conclusions. 
For example, the statement "Requirements are changed with time. We do not work in waterfall projects when a requirement is thought to be completed, cannot be changed, and then handed over to the developer. We have a parallel work in which we often realize that something was not expressed in a good way, it is often that we change the wording or more things a bit." was coded with the code "Changes during development". Similarly, the statement; "We mostly work with drafting the requirements during the sprint as they are not complete when we bring them into the sprint, so a part of our task is to make an investigation" were coded with the same code.

The coding resulted in synthesized themes organized in a document with related codes and key sentences. After the themes had been defined, they were used to draw conclusions on the appropriate level of abstraction to answer the research questions.

As thematic coding, with semantic equivalence partitioning, is subject to researcher bias, the first author validated the coding scheme with the second and third authors. 
We conducted this validation in the early stages of the analysis process. 
This was achieved by providing the second and third authors with an interview transcript and the codebook. 
The authors coded the corresponding transcript using the codebook. After the coding, the results were compared based on similarity. Results showed a high similarity: $74\%$ of the codes matched. We calculated this percentage as the number of sentences tagged with the same codes divided by the total number of coded sentences. This result was considered sufficient for the first author to proceed with the rest of the coding.

\section{Results}\label{sections:results}

This section presents the results of the interview study. First, we summarize how requirements are handled and utilized in Nordic Bank, and the perceived prevalence of bad requirements. The results that aim at answering research questions RQ1, R2, and RQ3 are presented in subsections~\ref{sections:resultsRQ1}, \ref{sections:resultsRQ2}, and \ref{sections:resultsRQ3}, respectively.

\subsection{Requirements Engineering Process at Nordic Bank}

This subsection presents the results that the participants provided in regard to the ways Nordic Bank utilizes requirements during the development process. We also include a brief analysis of the prevalence of bad requirements in the organization. 

\subsubsection{Workprocess}

\begin{figure}[tb]
\includegraphics[width=\textwidth]{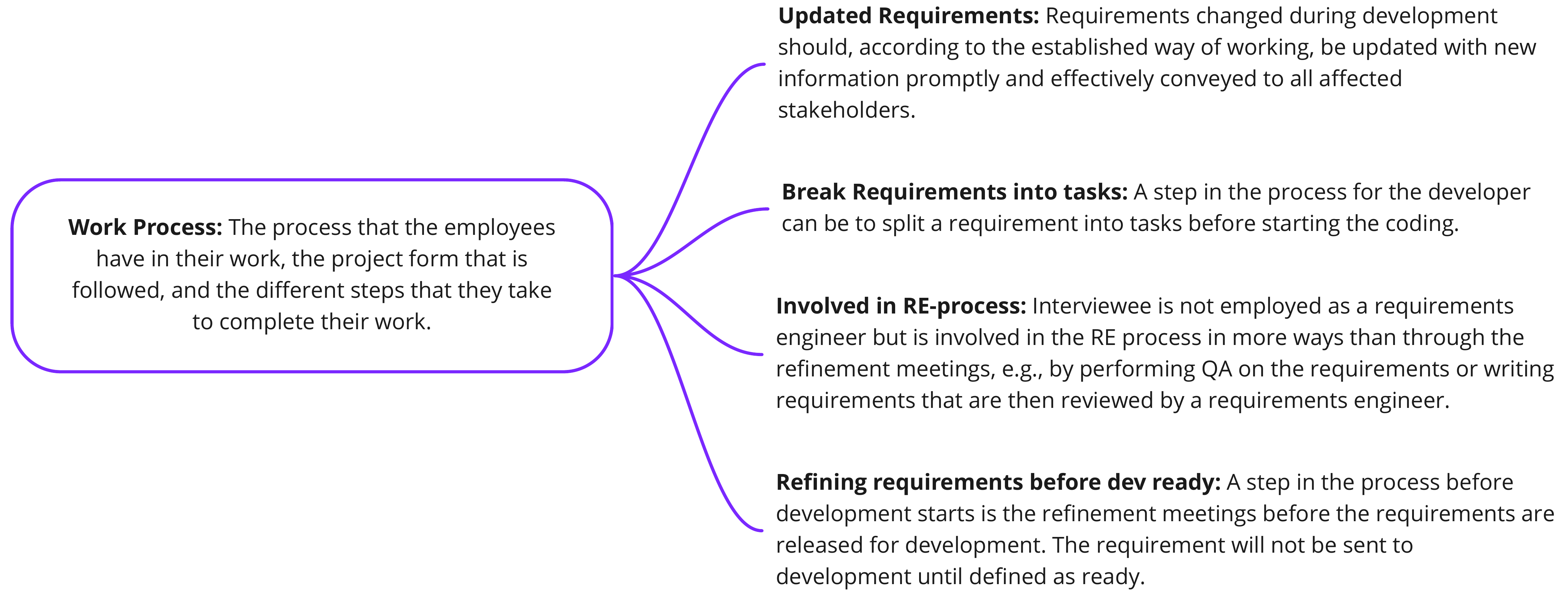}
\caption{Requirements work-process-related codes} \label{fig:workprocess}
\end{figure}

Fig.~\ref{fig:workprocess} presents the code book for the codes related to how requirements are handled during the development process. Four testers reported that they were involved in the requirements engineering process and performed requirements quality assurance activities. The developers also mentioned that within their teams, they carry out requirements refinement activities---Activities aimed at improving the content and understandability of the requirements---before the requirement reaches the status of \textit{ready-to-develop}. 

Participants also mentioned that it is possible to update the requirements once the development has started, but in those cases, all stakeholders should be informed about the change. 
Finally, it was also reported that it is a common practice to split requirements into smaller tasks to be carried out by the team or other teams.

\subsubsection{Bad Requirements Prevalence}

The testers involved in the quality assurance process for the requirements stated that the requirements, in general, in their understanding, were of good quality.
Additionally, they stated that they write preliminary test cases for all requirements, including edge case tests. 
Despite these efforts, they still discover many defects when they test the implementation. 
One interviewee stated, ``I think the quality is quite high, but that does not mean that if the requirements are good, the developers won’t make mistakes and bugs […] It would seem that we have bad user stories if I say that I give back 75\% of the stories for fixing because I find bugs”.
The same tester thought that it was mostly the developers' own fault and not caused by bad requirements, “It is not that they don’t understand or that might not mean that the requirements are bad, but this is just that they might not read the story enough or maybe interpret something differently.” 

A statement from a tester that did not perform QA checks on requirements before handing them to developers stated,``It’s important that I as a tester and the developers both understand it [the requirement] as it is written” when describing good requirements. 
Another interviewee stated, regarding requirements quality, ``If you are new in an area and in your role, then there is a higher demand on the quality of the requirements.'' 
Another tester involved with the RE process experienced that defects, in one-third to half of the times, were caused by bad requirements “It’s like 50-50. It doesn’t need to be bad requirements; it might be that some developer has missed something. But maybe less than 50\%, maybe 30\% are bad requirements''. 
However, because the actual requirements were not analyzed in the study, the testers' experiences are still unverified. 

The developers, testers, and requirements engineers agreed upon one point: refinement meetings are part of their RE-process. Still, they estimated the frequency of bad requirements to be about 10-25\% of the requirements.

\subsection{RQ1: How Software Development Practitioners Define Requirements Quality?}\label{sections:resultsRQ1}

\begin{figure}[tb]
\includegraphics[width=\textwidth]{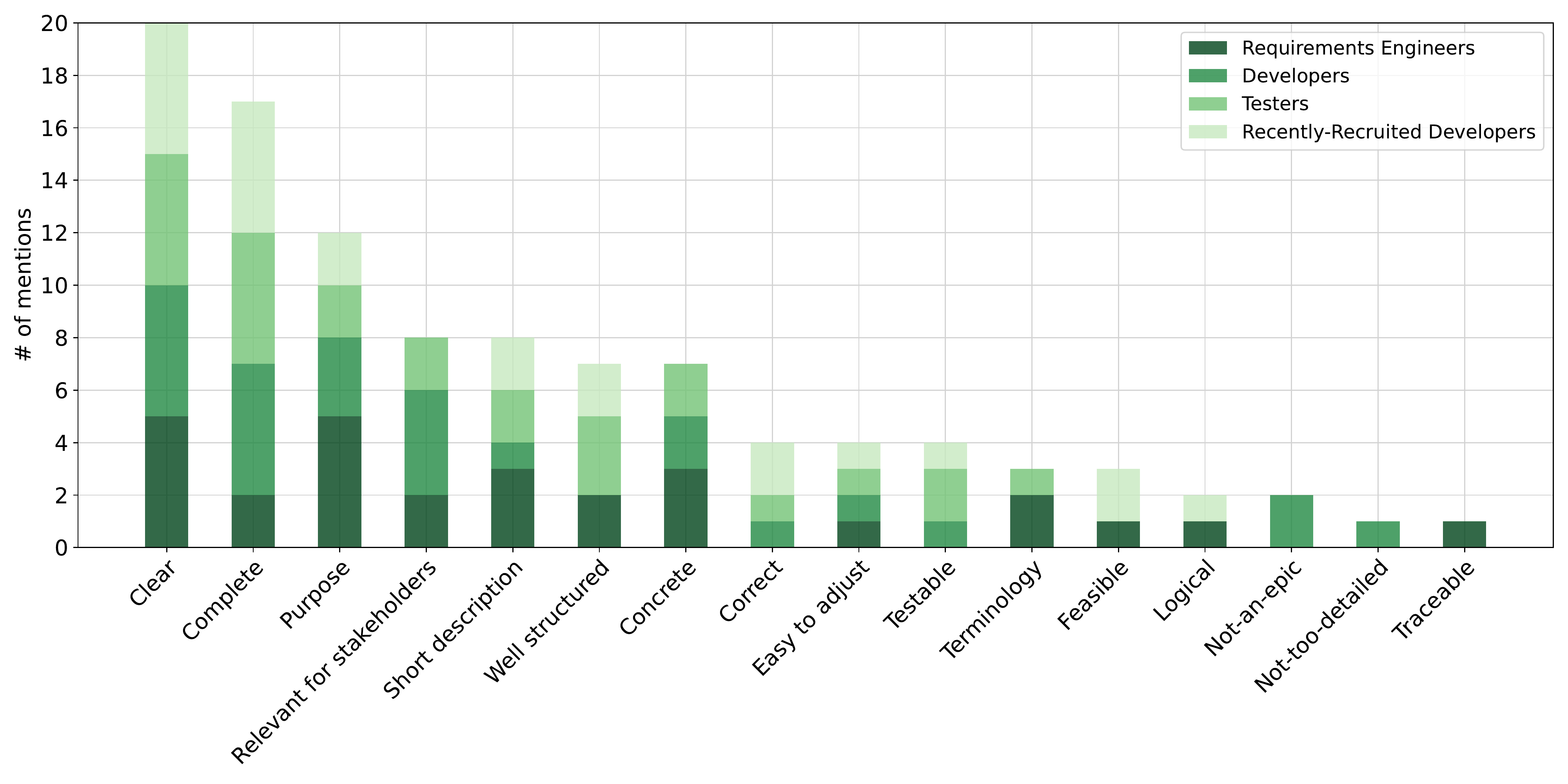}
\caption{Quality Characteristics of Good and Bad Requirements} \label{fig:qualities}
\end{figure}

Figure~\ref{fig:qualities} illustrates the requirements quality characteristics mentioned by the participants. 
The number of mentions accounts for each interviewee that mentioned each quality characteristic during the interviews. 

Every interviewee mentioned \textit{clear} as a characteristic of a good requirement. 
However, as clear might have multiple interpretations of what it means, which also can vary among roles, some interviewees might have mentioned this characteristic as a collective term for other characteristics, e.g., unambiguous and/or complete. Regardless, it seems that clarity, associated with ease of understanding, is considered a pivotal characteristic for all interviewees.
A tester said when describing good characteristics of requirements that “Clarity is, as mentioned, a keyword here, and there can be several different aspects that can make it [a requirement] clear.”

\begin{figure}[tb]
\includegraphics[width=\textwidth]{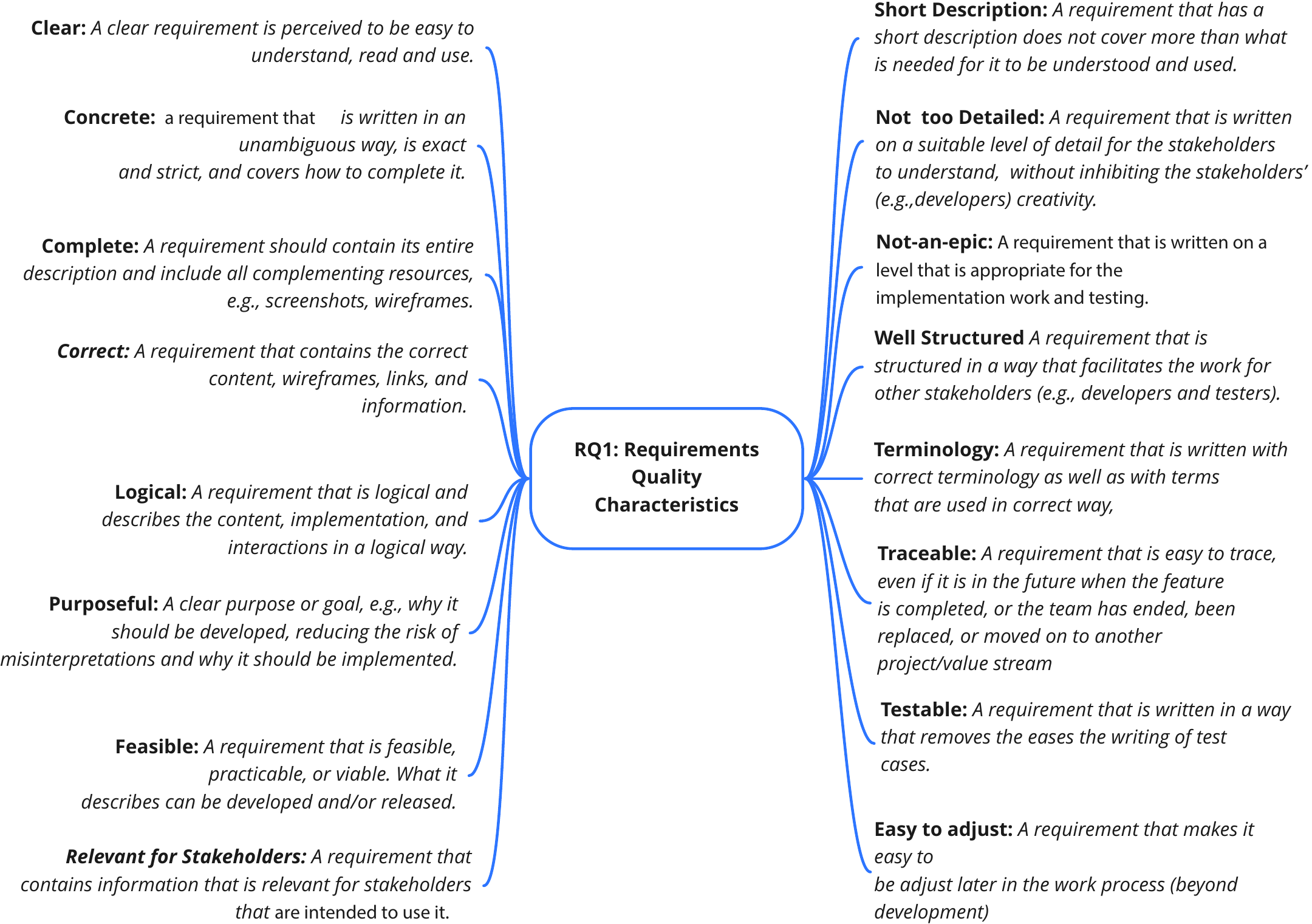}
\caption{Codebook: Characteristics of ``good'' requirements, with codes and descriptions of ``good'' requirements. Opposite descriptions would characterize ``bad'' requirements.}\label{fig:RQ1}
\end{figure}

The characteristic \textit{complete} was described by every developer, and most testers agreed that requirements have to be complete when they receive them. Some testers pointed out that changing a requirement while in testing would make the process much more complex. Only three requirements engineers explicitly mentioned complete as a characteristic of good requirements. However, one of those three requirements engineers pointed out that they accept, in their team, changing parts of requirements after developers receive them if they are aware of those likely changes. Fig.~\ref{fig:RQ1} presents the codebook with the description of the quality characteristics as discussed with the interviewees.

\subsection{RQ2: How does the perceived quality of requirements impact the work of practitioners of software development?}\label{sections:resultsRQ2}

\begin{figure}[tb]
\includegraphics[width=\textwidth]{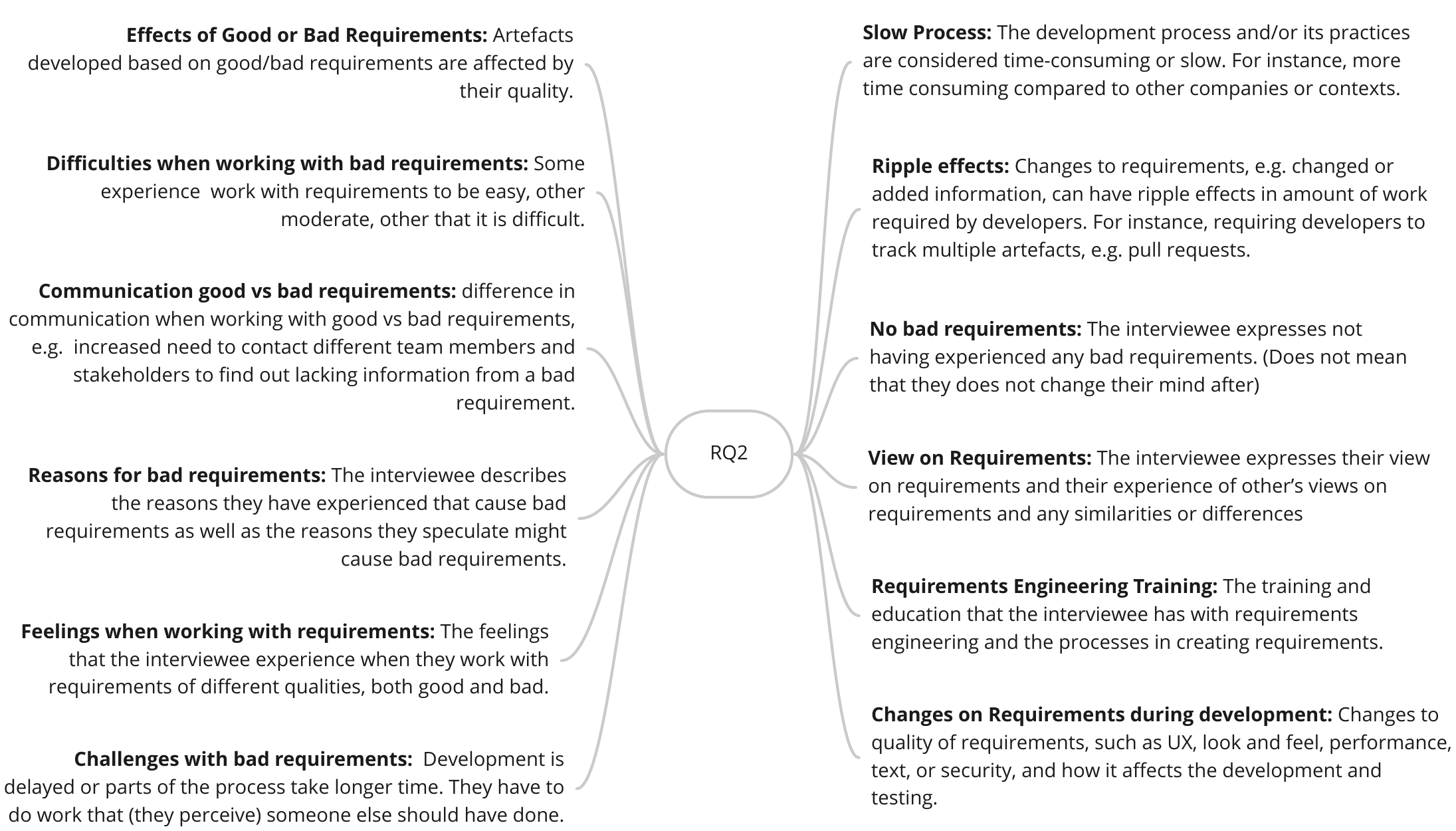}
\caption{Codebook: Codes Related to the impact of the quality of requirements} \label{fig:RQ2}
\end{figure}

In general, the participants experience that bad requirements cause delays in the development, that activities take longer time, and that they need to perform tasks that they perceive someone else should have done.
One possible cause of these experiences is the lack of communication within the team or with other teams or stakeholders. 
The interviewees also reported that good requirements positively impact in general, e.g., on code quality, shorten development time, or improve work satisfaction. 
In contrast, bad requirements have a negative impact on code quality, cost or work satisfaction. 
Fig.~\ref{fig:RQ2} presents the codebook with the main themes that emerged from the interview transcript analysis regarding RQ2.

\subsection{RQ3: What are the perceived causes and potential solutions of the poor quality of requirements?}\label{sections:resultsRQ3}

\begin{figure}[tb]
\includegraphics[width=\textwidth]{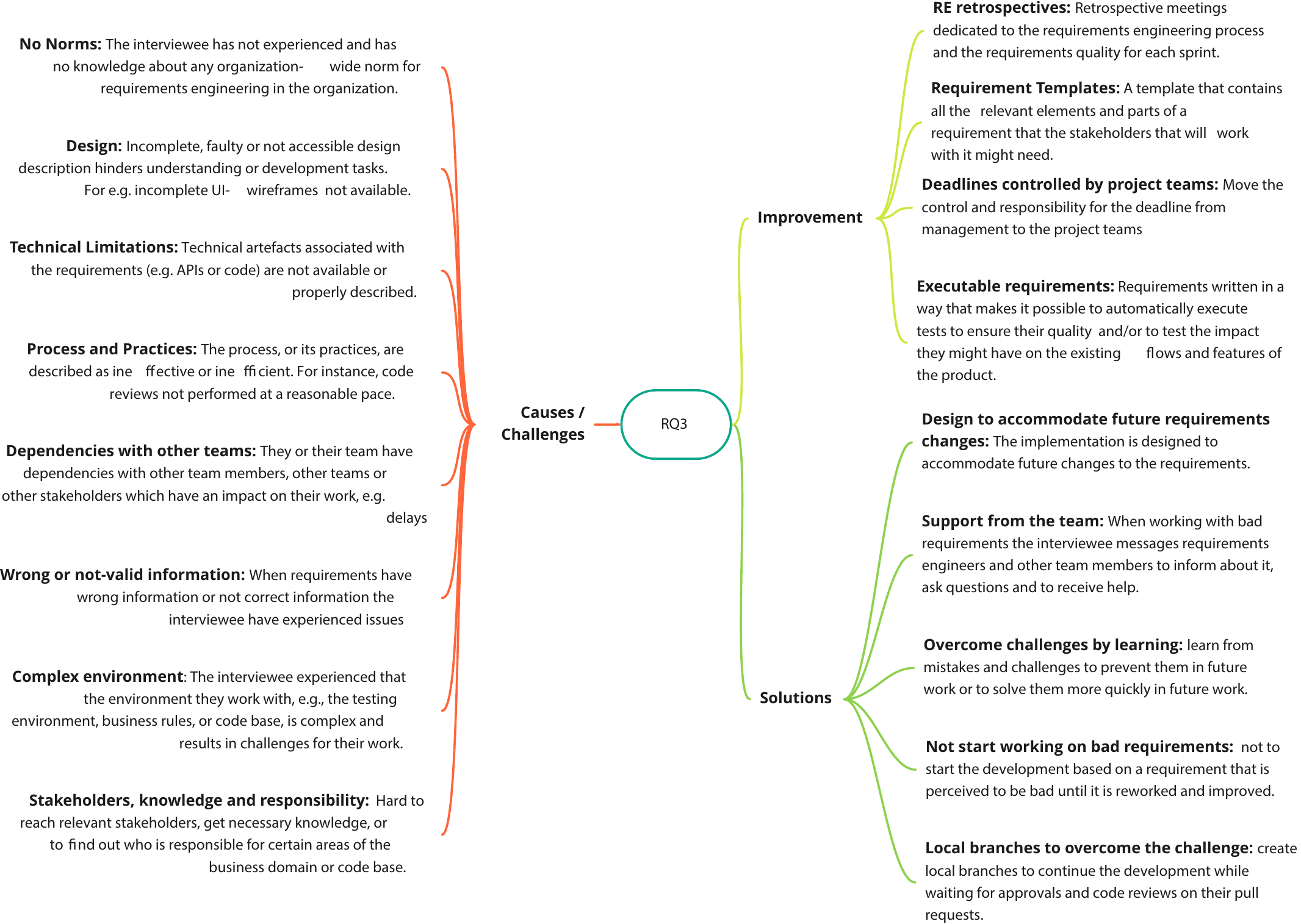}
\caption{Codebook: Codes related to causes, solutions and improvements for ``bad'' requirements} \label{fig:RQ3}
\end{figure}

Fig.~\ref{fig:RQ3} reports the code book with descriptions of the themes related to causes, challenges, potential solutions and process improvements to address low-quality requirements. 
The requirements engineers' most common suggestion for the cause of the poor requirements quality was that they got too tight deadlines that the team did not have any control over. 
Another plausible cause was stated to be a lack of agreement on what constitutes a good (or bad) requirement.

One of the improvements suggested by the interviewees was retrospective meetings for requirements engineers between different teams. 
Another alternative is to have some form of forum or other platforms for the requirements engineers to share knowledge, experience, and ideas and have workshops to share knowledge within the organization. 
Lastly, a suggestion was to move the responsibility and control over deadlines from managers to the development teams.

\section{Discussion}\label{sec:discussion}

\subsection{RQ1: How do software development practitioners define requirements quality?}

In this subsection, we discuss the main quality characteristics highlighted by the participants, mapping them to the quality characteristics included in the ISO~\cite{ISO29148} and IREB~\cite{IREB}.

Table~\ref{table:mapping} shows the mapping between the different characteristics highlighted by the participants to the ones in the standards.

\begin{table}[hbt!]
\begin{center}
\scriptsize
\caption{Mapping of the Characteristics from the Codebook to the ISO~\cite{ISO29148} and IREB~\cite{IREB} standards.}\label{table:mapping}

\begin{tabular}{p{0.25\textwidth} p{0.35\textwidth}p{0.4\textwidth}}
\toprule[1pt]
\bfseries\makecell[l]{Codes from\\ Codebook} 	& \bfseries\makecell[l]{ISO~\cite{ISO29148} Characteristics\\ \& Attributes} & \bfseries\makecell[l]{IREB~\cite{IREB} Characteristics\\ \& Attributes} \\
\midrule[1pt]
\textbf{Clear} 		& Unambiguous,  Comprehensible. & Unambiguous, Understandable	\\
\textbf{Concrete} & - & - \\
\textbf{Complete} 				& - 			& -												\\ 
\textbf{Correct}					& Correct 		& -												\\

\textbf{Purpose}					& Necessary 	& Necessary										\\
\textbf{Feasible}				& -				& -												\\

\textbf{Easy to adjust}			& -				& -												\\

\bfseries\makecell[l]{Relevant for\\ Stakeholders}          & Appropiate & -						\\
\textbf{Short Description}		& -				& Short and well-structured sentences			\\
\textbf{Not-too-detailed}		& -				& -												\\
\textbf{Well Structured} 		& -				& -												\\
\textbf{Terminology}				& Avoid open-ended non-verifiable terms, avoid subjective	& Defining and consistently using a  uniform terminology, avoid vague or, ambiguous terms and phrases						 \\

\textbf{Logical}					& -				& -												\\
\textbf{Traceable}				& -				& Traceable										\\
\textbf{Testable}				& Verifiable 	& Verifiable									\\

\textbf{Not-an-epic}				& -				& -												\\

\bottomrule[1pt]

\end{tabular}

\end{center}
\end{table}

As mentioned in Section~\ref{sections:resultsRQ1}, most developers and testers agree that a requirement has to be complete when they receive it. However, some testers do not necessarily agree that the requirements must be complete when the developers receive them. 
Requirements engineers argue that requirements are artefacts that can change during development. Testers do not mind if changes occur, but when the functionality is implemented and sent to them for testing, the requirements should be complete so that they can be verified.

One developer expressed that they worked in a team with a requirement standard that did not require the requirements to be complete when the developers received them. 
Instead, the norm in this team was that the requirements engineer would perform their own investigations while working on parts of the requirement that were considered to be  ``done''.
Similarly, other interviewees stated that working with incomplete requirements in their team was possible.

Some interviewees said that a requirement needs to give the stakeholders insight into the requirement's purpose and when it is considered done, i.e., the definition of done. 
Some quotes are ``The ones that make developers understand the scope,''. ``To understand the purpose of what we want to achieve''. Hence, not only should the requirements describe the functionality but also the acceptance criteria. 

One could argue that the requirements engineers should focus on domain knowledge and present requirements to the developers and testers that convey this knowledge. 
One participant expressed that the consequence of the lack of technical knowledge is that the requirements engineers do not know what is possible to implement and what is not, nor what limitations exist in one platform but not in another when they create features for both.

A subject that most interviewees brought up was that they perceived that a requirement should be written for the Stakeholders that will use it. ``It depends on the developer, the business analyst or the tester how they want it written''. ``In our team are several different roles that should understand it; our APO [agile product owner] looks at different parts than what our developer does''. Regarding how to deem a requirement, whether it is good or bad, a developer said: ``Maybe it’s not the ones that have the most insight in a user story [requirement] that should judge if it’s ready for development, but maybe those that have lesser knowledge in the area are the better persons to ask if it’s clear enough for them to understand what they are supposed to do.'' They base their subjective view on their technical and domain knowledge about the subject, which allows them to deal with assumptions more accurately.

\subsection{RQ2: How does the perceived quality of requirements impact the work of practitioners of software development?}

\subsubsection{Bad requirements require more communication}

Communication among requirements engineers, developers, testers, or other development teams, was the go-to practice for solving issues with the requirements for developers and testers. 
The logical outcome of this way of working is longer development time and an additional burden placed on the development team. 
A burden that could have been mitigated if the requirement had been better upfront. 
A developer stated, “Like I said earlier, a lot of the time that I want to put on programming is instead put on communication with BA:s [Business Analysts], UX [user experience] and other developers that have knowledge about what I’m working on. One does not develop so much; it is very much investigation work”.

Most interviewees, independent of role, have described that communication is different when working with good requirements compared to working with bad ones. 
One of the most mentioned differences is that more communication is needed within the team or with other stakeholders when working with a requirement perceived as bad. 
As stated by one of the interviewees, “Bad requirements generally result in that the time you could have put into making a good requirement from the beginning is instead spent on setting up meetings or communicate \textit{on the fly}.'' 

Another consequence of bad requirements commonly mentioned was ripple effects. As expressed by one of the requirements engineers: ``If I get questions [about a requirement] then I understand that I have forgotten something or haven’t been clear with something. That will make it take more time, there will be user stories [requirements] that stays, sprint after sprint, and will have to be brought back to me for further investigations.''

Another essential aspect is understanding of the area, domain and business rules for the requirements.
This includes all the work connected to them, directly affecting the team's communication. 
A statement from the interviews was: ``A lot of the understanding of a requirement is that you create a form of consensus. If you have a basic understanding of what needs to be done and how to solve it, then the requirements don’t need to be formulated in the same way as if you didn’t have it [the basic understanding].'' 

According to Zowghi and Nurmul~\cite{Zowghi2002}, the more often developers and customers communicate during the RE process, the less volatile the requirements become. 
In the organization where this study was performed, the requirements engineers acted as product owners (customers).

\subsubsection{Increased Workload and Doing Someone Else's Work}
Participants described that additional work with incorrect information resulted in having to redo the work after correcting the information. 
The quality of the requirement impacts this challenge, e.g. a clear, well-written and complete requirement might be easier to implement and test than one where the information is faulty.

Bad requirements can cause ripple effects in overhead communication, especially incomplete requirements cause such ripple effects and increased workload. 
As stated by the interviewees: “This slows us down. So, it’s kind of a joint work with the whole team, how we are coping with bad requirements, it’s kind of more work for the developers then of course for the testers.”, “Bad requirements create more work for us, testers because then we need to return to the business people, and the developers might have to rework something, so it leads to more work or rework.”

A severe consequence of receiving bad requirements is when a bad requirement is misunderstood by both the developer and tester, implemented, and then released. 
As expressed by one of the interviewees, “Then you realize that what came out in production is not correct, but we both [developer and tester] thought that it looked correct because ‘that is how it should work,’ then it can end up very wrong”. 
A requirements engineer said, “Sometimes one can write bad requirements, and it is developed according to them, and then you have to redo the work.” 

Four out of five requirements engineers have described those bad requirements they create also impact their work. 
The requirements engineer who did not experience an impact described that the questions did not impact them and that they viewed a part of their work as being prepared to explain the requirements to developers and testers. ``You can foresee that you might get questions and that you are the one that will get the bad feedback on the requirements.''

Interviewees experience not enough analysis performed on some of the requirements perceived as bad before handing them over to developers. 
In these cases, developers have to do work that they consider someone else should have done but which is required to overcome the lack in quality of the requirements. 
Most often, they need to perform investigations by going through documentation or contacting colleagues or other stakeholders, similar to what requirements engineers do when creating the requirements. 
The implications of this additional work are increased workload and work time in the development process, reduced work satisfaction and increased cost.
Additionally, sometimes, the developers find out that someone else has to perform work before they can do anything leading to dependencies and blocking of requirements. 

\subsection{Effects on Morale}

The Interviewees were asked to express how they feel when they work with bad and good requirements. 
We asked the requirements engineers to how it feels to work with creating requirements. Three of them mentioned that, when working with challenging requirements or receiving requirements from stakeholders that they perceive are bad, they feel frustration, stress, anger, and exhaustion. ``Exhausting is a word that comes to mind first.'' ``It can be anger and understanding that some people don’t have the same approach you have.''

Two requirements engineers did not mention receiving bad requirements. However, when describing how it felt when they created challenging requirements, they experienced it to be educational but also challenging in a positive way as they learn from the challenge, ``The challenge is in itself educational. I don’t see it as something negative but rather something that I learn from.'' 
One of the reasons for the big difference in the positive and negative experiences of challenges when creating requirements could be related to the type of challenge they encountered. 
The challenges that the two requirements engineers described a positive view and feelings from working with challenging requirements were:
\begin{enumerate*}[label=(\roman*)]
\item finding out stakeholders to contact for information;
\item challenges with some terminology;
\item how to get all the information from different sources together; and \item to write the requirements in a clear way for the developers and testers. 
\end{enumerate*}

The three requirements engineers that had a negative experience when working with challenging requirements described challenges such as: \begin{enumerate*}[label=(\roman*)]
\item technical debt, e.g., legacy code, that affected the creation of requirements,
\item that the test environment was perceived as unstable,
\item that some areas of the organization do not have any clear owner,
\item that legal and compliance aspects are difficult to work with, and
\item that the roadmap can be drastically changed without any heads-up by the managers.
\end{enumerate*}

When the requirements engineers are working with requirements they have made that the developers or testers perceive to be bad requirements, they describe that it feels sad and stressful, that it impacts their self-esteem, and that it drains their energy. ``It results in that I feel more stressed, I get less good work done, and it affects one’s own self-esteem.'' 

When working with requirements that they did not experience as challenging and when creating requirements that developers and testers perceived to be good, the requirements engineers describe it as fun and satisfying.

Most testers described that they experienced stress, frustration, a feeling of disappointment or dissatisfaction, and a loss of interest when they worked with bad requirements. 
As stated by two interviewees: ``When you have to talk with people, or when insecurity arises, you have to read it over and over again, you get frustrated. I lose interest if it is too bad''; ``Frustrating, and stressful. I might be maybe angry, or I might be disappointed.” 
They also explained that their work became less efficient and often resulted in more work because they had to communicate to other team members and often perform their tests again.

Feelings described by the interviewed developers when they received requirements they perceived as bad were: exhaustion, stress, frustration, a feeling of sadness, that they get feeling of doing something pointless, ineffective, and a waste of time, and that it was not fun when they worked with bad requirements. 
As stated by one interviewee: ``Facepalm [interviewee put their hand on their forehead], it’s a waste of time. Just a waste of time, resources, and energy”, “One gets sad and feels simply unproductive. You want the hours you put down on your work to be meaningful''.On the contrary, in general, the developers said they experienced positive feelings and that more work was done when they worked with good requirements.

One from each of the two developer groups, newly employed developers and developers that have worked for at least a few years, said they were not negatively emotionally affected by the bad requirements. 
One interviewee stated: ``Business as usual [laughing]. You can usually do something about bad requirements before you start working on them.'' 
Still, one of the interviewees described strong positive feelings when working with good requirements, “Effective. It’s stronger feelings when you have a clear good requirement that you can just work through”.

It is generally believed that morale has an impact on productivity. 
However, it has been difficult to prove in software engineering since both morale and productivity are difficult to measure~\cite{Storey2021,Weakliem2016}. 
Work morale can also affect a company’s attraction and retention of employees.
One possible consequence of lowered work morale could be that employees decide to leave the organization, ``Our developers are the ones that produce something of value, and if they are angry or sad over something possible to go live with, then we risk losing them. To onboard new developers is not a dance of roses''.
The quality of requirements that a requirements engineer work with might affect their and other Stakeholders’ morale. However, work morale can also be affected by more factors such as work environment or compensation.

\subsection{RQ3: What are the perceived causes and potential solutions of the poor quality of requirements?}
\subsubsection{Potential causes for poor quality requirements}
The organization has not adopted any norms for good requirements, and none of the interviewees had heard or experienced any form of such organization-wide norms or standards. 
However, several shared courses in the subject are available to the organization's employees. 
One requirements engineer said: ``I have covered the ones that come with the courses introduced in this company from different phases, such as ‘simplify,’ ‘SWAP,’ and now ‘SAFE’''. The closest to a shared norm that a few interviewees mentioned was 'SWAP' and the branching that derives from it, ``What we can lean on is 'SWAP' and the different branches that derive from it. But I wouldn’t say that there is any statement of ‘this is how you structure requirements at this company.’ It is more from team to team''.

A developer whose team requires that all requirements have to be complete before the developers receive them and that no changes are allowed, stated that they experienced that they still got incomplete requirements. 
He perceives a lack of knowledge or information by the requirements engineers.``Then we [the developers] need to step in and explain technically what is possible and what is not for the different platforms. It results in us more or less educating our requirements engineers.'' This statement highlights a possible root cause of the low quality of requirements, i.e., the requirements engineers lack technical knowledge.

\subsubsection{Potential Solutions and Improvements} Some interviewees had suggestions for how to solve the different causes for bad requirements that have arisen in the interviews. 
One of the long-term solutions is to enable more communication between requirements engineers across the organization in some form of platform.
One example given was to have retrospective meetings between requirements engineers, another to have a forum dedicated to RE, and a third would be to have a form of meeting for sharing knowledge, similar to tech talks that developers have.
As a complement to introducing a knowledge-sharing platform for requirements engineers, one could argue that a norm for creating requirements should be shared across the organization. 
This use of RE fora could be especially beneficial for organizations with several requirements engineers with different backgrounds and knowledge of requirements engineering, similar to the studied organization. 
One possibility is that such norms might naturally be developed and polished by the practitioners as a consequence of the activities and sharing of the knowledge-sharing platform.
Another suggestion from interviewees as a short-term solution was to move the control of the deadlines for projects that development teams worked on to the development teams. Thus, the requirements do not get rushed and enable a more agile way of working since fixed deadlines are more an aspect of the waterfall principles. 
However, one can speculate that such a change in the organization might be costly should the deadlines be moved forward repeatedly with delayed releases. 
Nevertheless, one can also argue that it can ease the quality assurance process. In addition, the practitioners would be less stressed and feel more work satisfaction. Hopefully, there should be fewer defects and less rework, possibly covering the extra cost that allows development teams to move deadlines forward might bring.

The expressed need for a platform dedicated to knowledge sharing between requirements engineers could indicate that the organization might benefit from an organization-wide norm for requirements engineering. 
However, many interviewees expressed a strong willingness to have flexibility and freedom in their work and their team’s ways of working. Therefore it should be considered to ensure that such a norm will be kept on a supporting level that does not encroach on the practitioners’ creativity.

Almost all interviewees, if not all, concur that if the organization experience more significant benefits to not having a standard norm, an improvement that the organization already should consider is a platform for knowledge sharing between requirements engineers. Imposing an organization-wide norm can hinder autonomy and heterogeneousness in the teams' ways of working.

There are several possibilities for how this platform could take shape. 
A straightforward example could be to have a forum dedicated to the RE-process and encouragement to the requirements engineers to use it and share the knowledge amongst themselves. Another example could be to have something similar to the tech talks that developers have, in which they have regular meetings to share news, knowledge, and insights. Some requirements engineers interviewed also asked for retrospectives for the requirements process, which they experienced a lack of.

The participants also suggested the use of templates and quality gates for requirements (i.e., not starting to work with low-quality requirements until they reach a certain quality level) as potential solutions to mitigate the effects of low-quality requirements.
\section{Threats to Validity}\label{sections:threats}

This section discusses threats to validity from four perspectives: construct validity, external validity, and reliability.

\noindent \textbf{Construct Validity} Construct validity is concerned with whether the studied measures reflect the constructs the researcher has in mind and what is stated in the research questions. 
The first author designed the flexible interview protocol and then reviewed it with the second and third authors. 
We acknowledge that the participants do not include all the relevant stakeholders in the organization. 
We tried mitigating this threat by involving participants with different roles and varying expertise from the companies.

\noindent \textbf{External Validity} concerns the extent to which the findings can be generalized outside of the studied case and whether they apply to other organizations. 
One of the misunderstandings about case study research is the inability to generalize from a single case~\cite{Flyvbjerg2006}. 
However, we have tried to build a theory to understand requirements quality, the impact of low-quality requirements, and causes and potential solutions by building analytic generalization through theories instead of gaining statistical generalizabilty.  
We have provided the characteristics of the case under analysis to allow us to evaluate its generalizability. However, still, further replications are needed to verify the results.

\noindent \textbf{Reliability} concerns whether the data and analysis are independent of the researchers. To increase the reliability, the second and third authors validated the coding scheme and the coding process by independently coding an interview transcript. 
The results of this independent coding matched for 74\% of the codes.

\section{Conclusions and Further Work}\label{sec:conclusions}

In this paper, we have presented an interview study to analyze how:
\begin{enumerate*}[label=(\roman*)]
\item
 practitioners from different roles define good and bad requirements; 
\item how the quality of the requirements impacts their work; and
\item what might be the causes for poor quality requirements, as well as potential solutions and improvements.
\end{enumerate*}

The results regarding the quality characteristics for requirements show that, although all interviewees agree that requirements should be clear, there is a wide range of views regarding the need to work with complete requirements. 
The participants highlighted that, in general, they experienced negative emotions, more work, and overhead communication when they worked with requirements they perceived to be of low quality. 
The participants suggested Requirements Engineering retrospectives, the use of templates, and quality gates for requirements (i.e., not starting to work with low-quality requirements until they reach a certain quality level) as potential improvements and solutions for low-quality requirements. 
Participants also suggested creating a requirements engineering forum (or guild) to disseminate requirements engineering knowledge better.

The most relevant further work is the replication of this study in other organizations to verify the results. Our preliminary results highlight some improvement areas that could be explored through longitudinal case studies or action research. Examples of those areas are the effects of establishing a knowledge-sharing forum for requirements engineers in organizations; or evaluating the cost, risk, and benefits of moving the control of deadlines from management to development teams in agile software development companies. These research areas could bring relevant results for researchers and software development organizations.

\subsubsection{Acknowledgements} This research was supported by the KKS foundation through the SHADE KKS H\"{o}g project (Ref: 20170176) and through the KKS SERT Research Profile project (Ref. 2018010) Blekinge Institute of Technology.

%
%
%
%
%
\bibliographystyle{splncs04}
\bibliography{bibliography}
\end{document}